\documentclass[journal=jacsat,manuscript=article]{achemso}
\usepackage{ulem}
\usepackage{float}
\usepackage{graphicx} 
\usepackage{booktabs}
\usepackage{amsmath}
\usepackage[version=3]{mhchem} % Formula subscripts using \ce{}

\author{Bipasa Samanta}
\affiliation{Department of Chemistry, Indiana University, Bloomington, IN 47405-7102, USA}
\email{bsamanta@iu.edu}
\author{Alexandru B. Georgescu}
\affiliation{Department of Chemistry, Indiana University, Bloomington, IN 47405-7102, USA}
\email{georgesc@iu.edu}

\title[An \textsf{achemso} demo]
  {In-Plane Ni–O–Ni Bond Angles as Structural Fingerprints of Superconductivity in Layered Nickelates: Effects of Pressure, Strain, Layering, and Correlations}

\abbreviations{DFT+U, Superconductors, High Pressure, Electron Correlation, Dimer Effect}
\keywords{American Chemical Society, \LaTeX}

\begin{document}

%%%%%%%%%%%%%%%%%%%%%%%%%%%%%%%%%%%%%%%%%%%%%%%%%%%%%%%%%%%%%%%%%%%%%
%% The "Coventry" environment can be used to create an entry for the
%\begin{tocentry}

%This box and the associated title will always be printed on a
%separate page at the end of the document.

%\end{tocentry}

\begin{abstract}
We investigate the structural and electronic conditions conducive to superconductivity in layered nickelates using density functional theory with Hubbard corrections (DFT+$U$). For both the bilayer and 1–3 polymorphs of \ce{La3Ni2O7}, we find that the in-plane Ni–O–Ni bond angles under pressure strongly correlate with the experimentally observed superconducting transition temperature ($T_c$) dome, and may serve as a reasonable proxy. Under compressive strain, the bond angles straighten, peaking near 2\% strain—consistent with experimental reports of superconductivity in strained bilayer thin films. However, the bond angles at this strain are more bent than those achieved under hydrostatic pressure, correlating with a lower $T_c$.
We show that increasing the number of \ce{NiO2} layers, as in \ce{La4Ni3O10}, or substituting heavier rare-earth elements (e.g., Pr) raises the pressure required to reach the structural configuration associated with superconductivity. Our results indicate that these systems require higher external pressure to achieve in-plane bond straightening.
Varying the on-site Coulomb interaction $U$ reveals that stronger electronic correlations delay the structural transition and favor high-spin states. This suggests that moderate correlation strength may be optimal for superconductivity, with stronger correlation preventing the formation of favorable bond geometries.
Electronic structure analysis shows that the Ni $e_g$ orbitals dominate near the Fermi level and shift downward with pressure, enhancing Ni–O hybridization. These results highlight how pressure and strain tune structural features that may be essential for engineering high-$T_c$ phases in nickelate superconductors.

\end{abstract}

\section{Introduction}
Since the discovery of superconductivity in nickelates, first reported in Nd$_{0.8}$Sr$_{0.2}$NiO$_{2}$ thin films with a superconducting transition temperature ($T_c$) of 15 K \cite{li2019superconductivity}, there has been renewed interest in the study of nickelate superconductivity. This breakthrough sparked further reports of superconductivity in the RNiO$_2$ family \cite{osada2020superconducting, osada2021nickelate, zeng2022superconductivity} and in the quintuple-layer compound \ce{Nd6Ni5O12} \cite{pan2022superconductivity}. The initial focus on nickelates was driven by their structural and electronic resemblance to cuprates \cite{botana2020similarities}: both contain transition-metal atoms in square planar coordination with oxygen, mimicking the CuO$_2$ planes in cuprates, and are at or near a $d^9$ electronic configuration.

\begin{figure}[h]
    \centering
    \includegraphics[width=\textwidth]{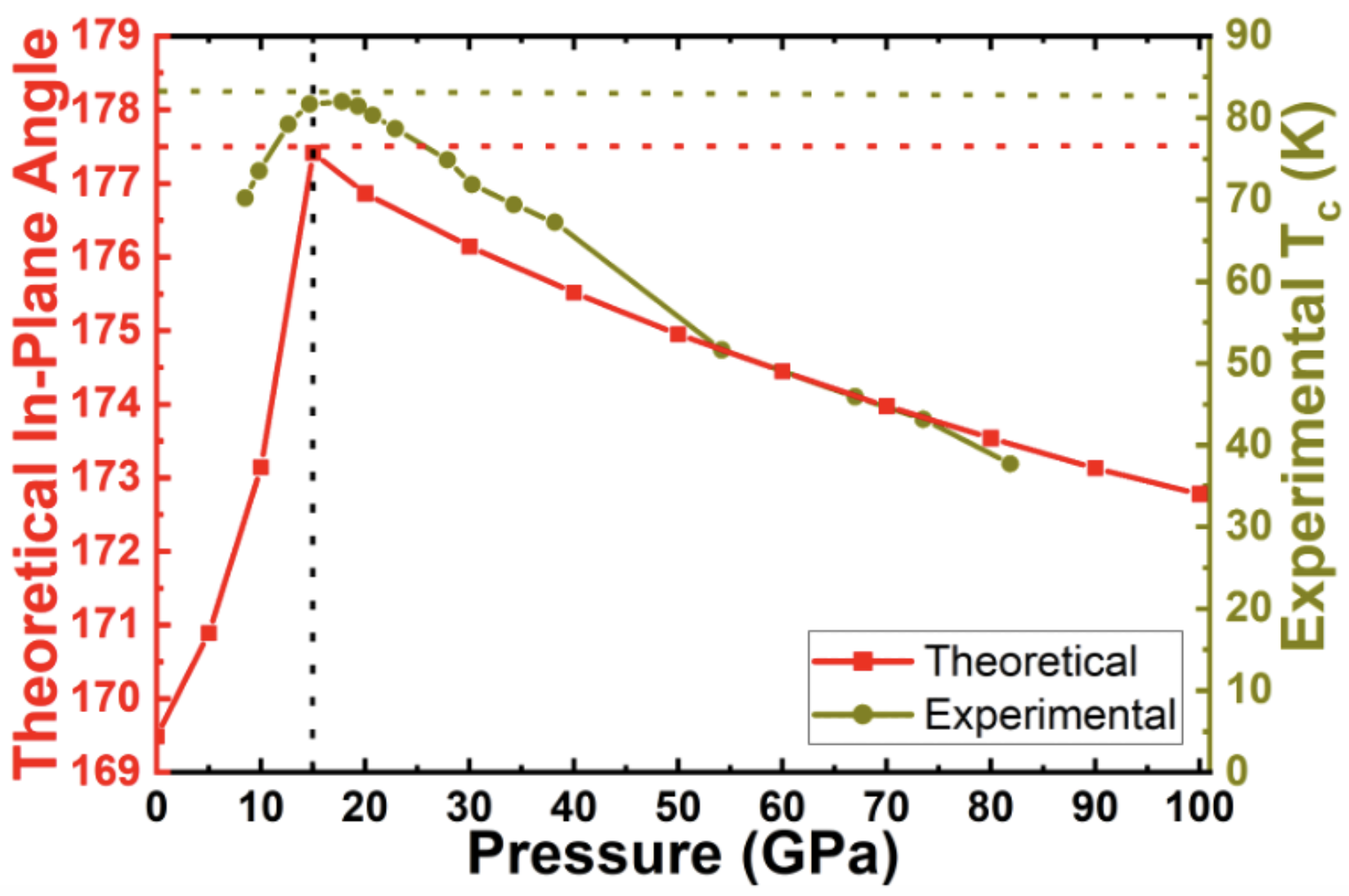} 
   \caption{Strong correspondence between the theoretically obtained in-plane Ni-O-Ni bond angles in the bilayer \ce{La3Ni2O7} at U=0 and the experimentally measured T$_c$. Experimental data digitized from Ref \citep{li2025identificationsuperconductivitybilayernickelate}}
    \label{fig:comparison}
\end{figure}

Despite these similarities, key differences exist. Unlike cuprates, which require hole doping to become superconducting, nickelates can exhibit superconductivity without it \cite{parzyck2024superconductivityparentinfinitelayernickelate}. Moreover, their $T_c$ values are significantly lower ($\sim$15 K in nickelates vs. 80–160 K in cuprates) \cite{li2019superconductivity, osada2020superconducting, osada2021nickelate, zeng2022superconductivity, osada2020phase}. Achieving superconductivity in nickelates required stabilizing Ni in the +1 oxidation state, which was accomplished by removing apical oxygen atoms in epitaxially constrained thin films \cite{li2019superconductivity, li2020superconducting}.

A new direction arose in 2023 with the discovery of superconductivity in the bulk Ruddlesden–Popper perovskite nickelate \ce{La3Ni2O7}, which exhibited a much higher $T_c$ of ~80 K under high pressure \cite{sun2023signatures} comparable to that of cuprates. This system preserves the NiO$_6$ octahedral framework without requiring apical oxygen removal or chemical doping. Superconductivity was induced under moderate hydrostatic pressure (14.0-43.5 GPa), prompting extensive studies to understand the mechanism behind this behavior \cite{zhang2024high, zhou2025investigationskeyissuesreproducibility, you2025unlikelihood, PhysRevX.14.011040, wang2024structure}.

Unlike cuprates (Cu$^{2+}$, $d^9$ configuration), Ni in \ce{La3Ni2O7} has an average +2.5 oxidation state, corresponding to 7.5 electrons in the \textit{d} orbital \cite{yamaura1999la3ni2o7, tofield1992electronic, liu2023evidence, cai2024low, chen2025charge}. This distinction has motivated a surge in theoretical studies exploring the electronic structure of \ce{La3Ni2O7} under pressure \cite{lechermann2023electronic, christiansson2023correlated, liao2023electron, labollita2024electronic}. These works examine how pressure modifies orbital hybridization, strengthens electron correlations, and shifts the relative positions of Ni $3d_{z^2}$ and $3d_{x^2 - y^2}$ orbitals near the Fermi level, affecting the material’s superconducting properties.

Recent studies have investigated \ce{La3Ni2O7} using diverse theoretical methods, including DFT+$U$, DFT+DMFT, GW+DMFT, and model Hamiltonians \cite{Shen_2023, yang2024orbital, zhang2023electronic, lu2024interplay, lu2024interlayer, xie2024strong, wang2024structure, Sakakibara_2024, Yang_2023, gu2023effectivemodelpairingtendency, Lechermann_2023, lu2025spin, W__2024, Tian_2024, wu2024superexchange}. These works emphasize the enhanced orbital hybridization, intra- and inter-layer hopping, and superexchange interactions under pressure, all of which are thought to support superconductivity.

Following the discovery of superconductivity in \ce{La3Ni2O7} with Ni in a +2.5 oxidation state, the research focus has largely centered on local $d$-orbital physics, superexchange mechanisms, and $s^{\substack{+ \\ -}}$ -wave pairing, seeking parallels to cuprates \cite{lu2024interplay, chen2025charge, green2016bond, xie2024strong, Liu_2023}. Our focus here is more specifically on the structural parameters and their evolution under different conditions.

In this work, we show that Ni-O-Ni in-plane bonds as a function of pressure are strongly correlated to the experimentally measured Tc. We also find that, for an optimal Ni-O-Ni bond straightening we need:
\begin{itemize}
    \item At 0 strain, optimal pressure: approximately 15~GPa for bilayer \ce{La3Ni2O7}.
    \item optimal compressive strain: around 2\% using SLAO substrate for \ce{La3Ni2O7} without additional pressure - in agreement with recent experiment \citep{strain}.
    \item Low number of perovskite layers: increasing the number of layers increases the pressure at which the bonds straighten
    \item La as the A-site cation: Pr for example bends the bond angles, leading to a higher optimum pressure.
    \item Low Hubbard \( U \): high \( U \) increases the energy cost for the low-spin state at straight bond angles. High \( U \) leads to a stronger preference for a bent-bond high-spin state. While electron correlations may be key to superconductivity, strong electron correlations are in competition with the structural transition empirically required for superconductivity to be observed.
    \item Non-magnetic or ferromagnetic/paramagnetic states: antiferromagnetic (AFM) ordering leads to an increase in the pressure at which the structural transition occurs. We note that in rare earth nickelates, a ferromagnetic state is often used as a proxy for the paramagnetic state\citep{stoica2022magnetic}.
\end{itemize}

 We systematically investigated how structural parameters evolve under pressure and strain and how these are influenced by electron correlation (modeled using DFT+$U$), magnetic ordering, rare-earth substitution (Pr for La), and the number of NiO$_2$ perovskite layers. We find that at the superconducting transition pressure, the out-of-plane Ni–O–Ni bond angle becomes 180°, while the in-plane angle reaches a maximum and then decreases, forming a structural dome akin to the superconducting dome. This transition coincides with a symmetry change from orthorhombic (\textit{Amam}) to tetragonal (\textit{I4/mmm}). The pressure required for this phase change increases with stronger electron correlation, a larger number of perovskite layers, and substitution with heavier rare-earth elements.

We further show that magnetic ordering affects the transition pressure and structural parameters. Specifically, magnetic order stabilizes distorted geometries, raising the pressure needed to achieve the tetragonal phase and out-of-plane bond alignment. Overall, our results reveal a complex interplay between structure, pressure, electron correlation strength, and magnetism in determining the emergence of superconductivity in layered nickelates.

\section{Results and discussion}
\subsection{Structural Analysis}

\begin{figure}[H]
    \centering
    \includegraphics[width=\textwidth]{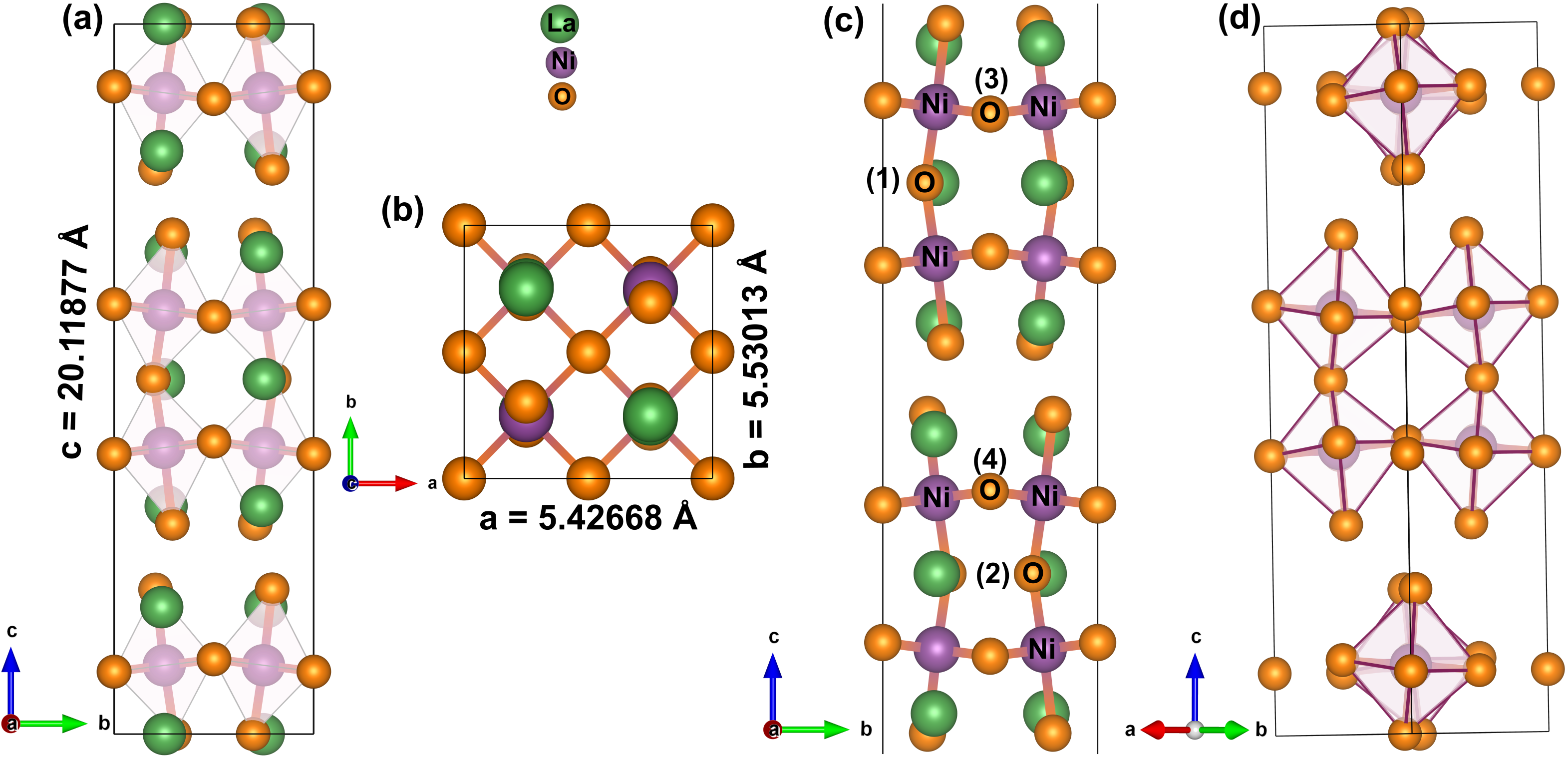} 
   \caption{(a) Front view of the unit cell of bulk bilayer \ce{La3Ni2O7}. (b) Top view showing the xy plane. Green, violet, and golden spheres represent La, Ni, and O atoms, respectively. (c) A doubled unit cell along the c-axis shows the two bilayers separated by a rock-salt layer, with four Ni–O–Ni bond angles labeled: Ni–O–Ni(1) and Ni–O–Ni(2) are out-of-plane angles, while Ni–O–Ni(3) and Ni–O–Ni(4) are in-plane. (d) NiO$_6$ octahedral tilting in the relaxed structure (La atoms are omitted for clarity).}
    \label{fig:example1}
\end{figure}

To elucidate structural signatures associated with superconductivity in \ce{La3Ni2O7}, we performed first-principles calculations under hydrostatic pressure. Starting with the ferromagnetic (FM) configuration and \textit{U}=0 eV, we relaxed the structure using variable-cell optimization. The optimized orthorhombic structure (space group \textit{Amam}) at ambient pressure is shown in Figure~\ref{fig:example1}a, with relaxed lattice constants of a = 5.42668 \AA, b = 5.53013 \AA, c = 20.11877 \AA, and $\alpha = \beta = \gamma = 90^\circ$. At zero pressure, the NiO$_6$ octahedra exhibit tilting (Figure~\ref{fig:example1}d), and all eight Ni atoms possess identical magnetic moments of approximately 0.34 $\mu_B$ in the FM ground state.

\begin{figure}[H]
    \centering
    \includegraphics[width=0.6\textwidth]{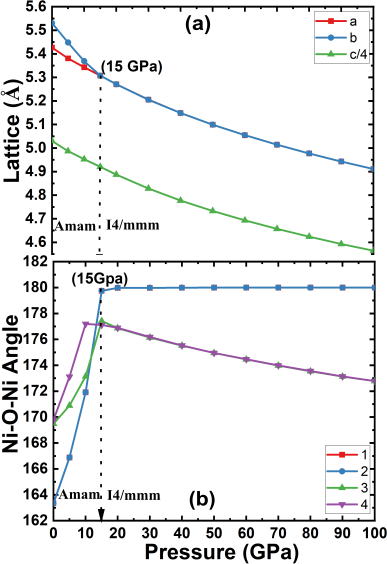} 
    \caption{(a) and (b) show how increasing pressure affects the a, b, and c lattice constants and the four Ni–O–Ni bond angles in \ce{La3Ni2O7} (as described in Figure 1d). The red line in (b) overlaps with the blue line.}
    \label{fig:example2}
\end{figure}

External pressure ranging from 5 to 100 GPa were applied to examine their influence on the structural and electronic properties. All three lattice parameters decrease monotonically under pressure, indicating isotropic compression. A structural phase transition occurs around 15 GPa, where the symmetry changes from orthorhombic (\( a \neq b \neq c\)) to tetragonal (\( a = b \neq c\)), with a corresponding space group change from \textit{Amam} to \textit{I4/mmm} (Figure~\ref{fig:example2}a). This transition, consistent with experimental observations by Sun et al. \cite{sun2023signatures}, confirms that our model captures key structural features under pressure.

The structural transformation is closely associated with changes in Ni–O–Ni bond angles. The out-of-plane Ni–O–Ni angle increases from 168° at ambient pressure to 180° near 15 GPa and remains fixed thereafter (Figure~\ref{fig:example2}b, red and blue lines; the red line overlaps with the blue due to angle convergence). In contrast, the in-plane Ni–O–Ni angle increases from approximately 169° to a maximum of 177° at the transition pressure ($\sim$ 15\,\text{GPa}) before decreasing at higher pressures (Figure~\ref{fig:example2}b, green and violet lines). This behavior correlates with the pressure dependence of the superconducting transition temperature T$_c$, resembling the transport data of Li et al. \cite{li2025identificationsuperconductivitybilayernickelate}, where a maximum T$_c$ of 83 K is observed. Similar dome-shaped superconducting behavior is observed in cuprates, where doping induces a first-order structural transition~\cite{timusk1999pseudogap}.

Concomitant with the transition, NiO$_6$ octahedral tilting is progressively suppressed with increasing pressure and vanishes entirely near $\sim$ 15\,\text{GPa}, resulting in fully aligned octahedra. Simultaneously, Ni–O–Ni bond lengths decrease from $\sim$ 3.9 \AA at 0 GPa to $\sim${} 3.6 \AA at $\sim$ 15\,\text{GPa} (transition pressure) due to lattice compression. These results suggest that the superconducting phase of \ce{La3Ni2O7} emerges from a compressed, symmetric tetragonal structure, defined by linear Ni–O–Ni bonds, shorter bond lengths, and the absence of octahedral tilting - the structural features stabilized at elevated pressures. 

\subsection{Electron Correlation Effect (U)}

To evaluate the role of electron correlations in \ce{La3Ni2O7}, we conducted DFT+\textit{U} calculations using a ferromagnetic (FM) configuration, systematically varying the on-site Coulomb interaction parameter \textit{U} from 0 to 5 eV. At \textit{U} = 0 eV, the magnetic moment per Ni atom decreases slightly from 0.34 to 0.31~$\mu_B$ across the transition pressure ($\sim$15 GPa). While prior studies have reported a pressure-induced spin-state transition from high-spin to low-spin configurations near this critical point, coinciding with the onset of superconductivity\cite{christiansson2023correlated, labollita2024electronicstructuremagneticproperties}—our calculations at \textit{U} = 0 eV do not capture such a transition.

Although the \textit{U} = 0 eV approximation reproduces the experimental lattice parameters and transition pressure with reasonable accuracy, it may not accurately capture the effects of correlations on the structure. Nickel, with its partially filled and spatially localized 3\textit{d} orbitals, exhibits significant on-site electron-electron interactions that are inadequately described by conventional DFT. To address this limitation, we employ the DFT+\textit{U} framework, where the Hubbard \textit{U} term explicitly accounts for on-site Coulomb repulsion. This correction is assumed to provide a better description of localized electronic states and their influence on both structural and magnetic properties in \ce{La3Ni2O7}.

\begin{figure}[H]
    \centering
    \includegraphics[width=0.9\textwidth]{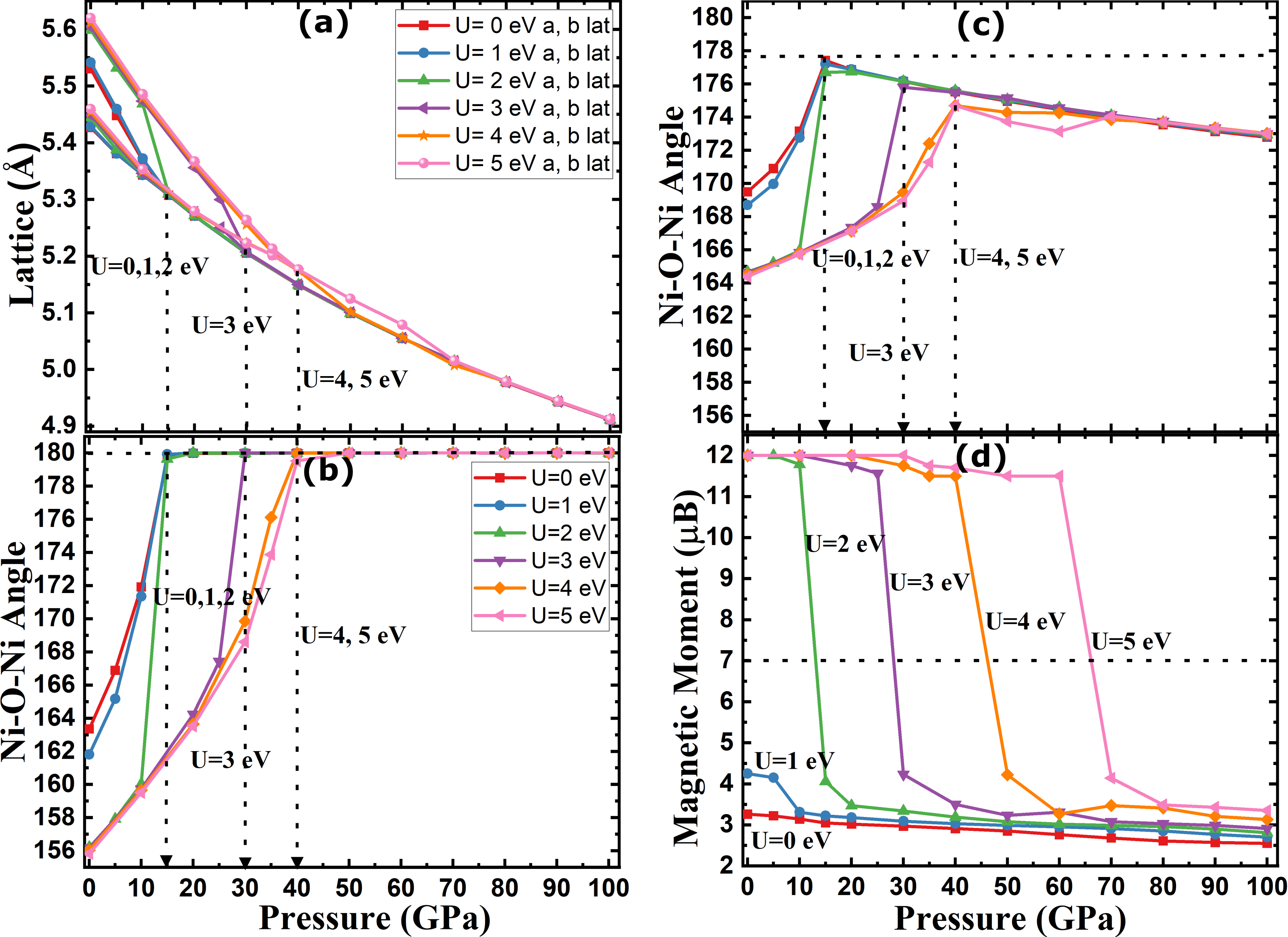} 
    \caption{Change in the \ce{La3Ni2O7 } structure in terms of (a) a and b lattice constant (b) out-of-plane Ni-O-Ni bond angle (c) in-plane Ni-O-Ni bond angle and (d) magnetic moment as the effect of increasing \textit{U} and pressure. The legends for figure (b) hold for (c) and (d) as well.}
    \label{fig:example3}
\end{figure}

The influence of \textit{U} on structural and magnetic responses is summarized in Figure~\ref{fig:example3}. As shown in Figure~\ref{fig:example3}a, increasing \textit{U} shifts the structural transition (defined by the condition $a=b$, and out-of-plane Ni-O-Ni bond angle aligned) to higher pressures. This shift reflects enhanced electron localization with larger \textit{U}, which stabilizes high-spin states and is energetically less favorable to straighten the Ni–O–Ni bond angles to 180$^\circ$. As a result, higher pressures are required to overcome this localization and drive the transition to the tetragonal phase. Figure~\ref{fig:example3}b corroborates this trend, showing a delayed onset of the ideal 180$^\circ$ Ni–O–Ni bond angle with increasing \textit{U}.

The evolution of the in-plane Ni–O–Ni bond angle with pressure is shown in Figure~\ref{fig:example3}c. For all values of \textit{U}, the bond angle increases with pressure, reaches a maximum near the structural transition point, and then decreases at higher pressures, exhibiting a dome-like trend reminiscent of the superconducting phase diagram, which is repeated across all $U$ values. Notably, as \textit{U} increases, the peak shifts to higher pressures and lower maximum bond angles, indicating that stronger correlations enhance Ni–O–Ni bending. This behavior suppresses bond straightening and electron delocalization - two features commonly associated with the emergence of superconductivity.

 Figure~\ref{fig:example3}d presents the pressure dependence of the magnetic moment. At \textit{U} = 0 and 1 eV, the moment remains nearly unchanged, showing no evidence of a spin-state transition. At \textit{U} = 2 eV, a pronounced reduction in magnetic moment occurs near the structural transition pressure, signaling a high-spin to low-spin transition. For \textit{U} = 3–5 eV, the spin-state transition is observed only after the structural transition has occurred, and the associated critical pressures are substantially overestimated. Moreover, large \textit{U} values tend to over-localize electrons, thereby favoring the high-spin state and impeding the electron delocalization necessary for superconductivity.

Among the tested values, \textit{U} = 2 eV yields the best agreement with experimental observations, capturing both the structural phase transition and the magnetic moment collapse near the experimentally reported transition pressure of $\sim$15 GPa. These results highlight a delicate interplay between electron correlation and pressure-driven structural changes: while increasing \textit{U} promotes spin localization, external pressure favors structural symmetry and electronic delocalization. At \textit{U} = 2 eV, these competing effects are balanced. This highlights the importance of 'just right' electron correlation strength: while correlations may play a key part in the superconductivity in these materials, correlations that are too strong prevent the structural transition needed for superconductivity to occur.

\subsection{Antiferromagnetic Ordering Effect}
\begin{figure}[H]
    \centering
    \includegraphics[width=0.5\textwidth]{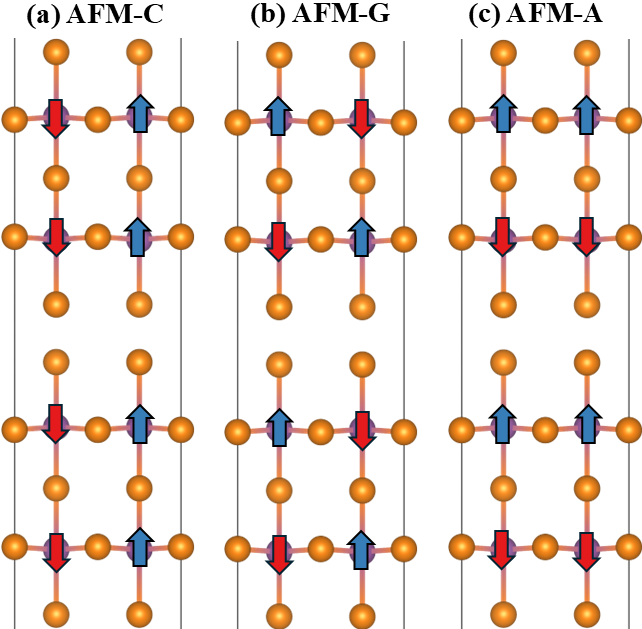} 
    \caption{Magnetic configurations explored. Up and down spin are represented by blue and red arrows.  (a) AFM-C (b) AFM-G (c)AFM-A}
    \label{fig:example4}
\end{figure}

The inclusion of on-site Coulomb interactions (\textit{U}) influences both magnetic ordering and the structural transition pressure in \ce{La3Ni2O7}. To further investigate the interplay between electron correlation and magnetic order, we examined three representative antiferromagnetic (AFM) configurations - AFM-C, AFM-A, and AFM-G, as illustrated in Figure~\ref{fig:example4}. In the AFM-C configuration, Ni spins alternate along adjacent columns; in AFM-A, spins alternate along rows; and in AFM-G, a checkerboard arrangement ensures that each Ni atom is surrounded by neighbors with opposite spins.

\begin{figure}[H]
    \centering
    \includegraphics[width=\textwidth]{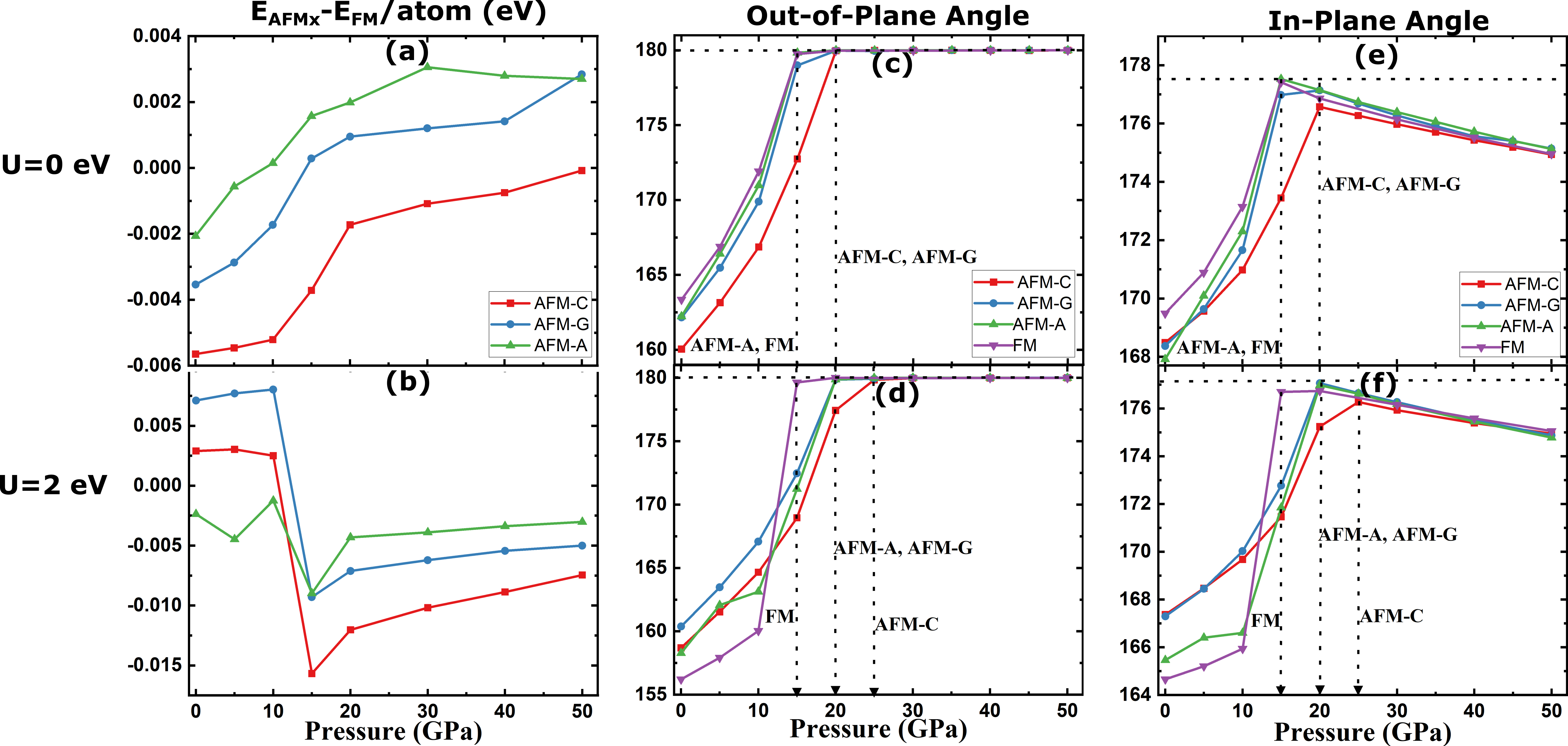} 
    \caption{(a) and (b) show the energy difference of the optimized AFM structures of $La_3Ni_2O_7$ with respect to FM structure. $E_{AFMx}$ refers to the energy of  optimised structure where x represents \textit {C,G and A}. $E_{FM}$ refers to the energy of the ferromagnetic structure. (c) and (d) represents the variation of the out-of-plane Ni-O-Ni bond angle, and (e) and (f) represent the in-plane Ni-O-Ni angle variation with increasing pressure.}
    \label{fig:example5}
\end{figure}

We evaluated the relative stabilities of these AFM configurations against the ferromagnetic (FM) ground state at \textit{U} = 0 and 2 eV. In general, AFM configurations promote stronger electron correlations, leading to increased electron localization and narrower bandwidths. Figures~\ref{fig:example5}(a–f) summarize the effects of magnetic ordering and electron correlation on the energetics and the structural parameters as a function of applied pressure.

Figures~\ref{fig:example5}a and ~\ref{fig:example5}b display the energy differences between the AFM and FM configurations as a function of pressure. At \textit{U} = 0 eV, AFM-C is the most stable among the antiferromagnetic configurations, followed by AFM-G and then AFM-A. When \textit{U} is increased to 2 eV, AFM-A becomes the most stable AFM phase at low pressures. However, as pressure increases, its stability decreases relative to AFM-C, which becomes the lowest-energy AFM state above $\sim$15 GPa. This pressure-induced crossover in stability reflects a reordering of magnetic phases and is consistent with the trend observed at \textit{U} = 0 eV.

Figures~\ref{fig:example5}c and ~\ref{fig:example5}d present the pressure dependence of the out-of-plane Ni–O–Ni bond angles for each magnetic configuration. At \textit{U} = 0 eV, both the FM and AFM-A configurations exhibit structural transitions near 15 GPa, whereas AFM-C and AFM-G require higher pressures ($\sim$20 GPa). For \textit{U} = 2 eV, only the FM state reproduces the experimentally observed transition pressure of $\sim$15 GPa. The inclusion of stronger electron correlations systematically shifts the transition pressure to higher values: approximately $\sim$20 GPa for AFM-A and AFM-G, and $\sim$25 GPa for AFM-C. Across all cases, AFM-C consistently exhibits the highest transition pressure.

Figures ~\ref{fig:example5}e and ~\ref{fig:example5}f illustrate the evolution of the in-plane Ni–O–Ni bond angle with pressure. Regardless of magnetic ordering, the bond angle initially increases, peaks near the transition pressure, and then decreases, exhibiting a dome-like behavior reminiscent of the "superconducting dome" seen in pressure–temperature phase diagrams. As electron correlation increases (via higher \textit{U} or AFM ordering), the dome shifts to higher pressures with reduced peak bond angles, indicating increased structural distortion and suppressed bond linearization, which are caused by stronger correlations.

In summary, our results reveal that both electron correlation and magnetic ordering play a decisive role in the pressure-driven structural evolution of \ce{La3Ni2O7}. The AFM configurations exhibit distinct stability regimes under pressure, with AFM-C consistently requiring the highest transition pressure. The inclusion of a finite \textit{U} not only modifies the energetic hierarchy among magnetic states but also delays the onset of the structural phase transition. These findings underscore the importance of simultaneously accounting for magnetic ordering and electron correlation when modeling the pressure-dependent behavior of correlated oxide systems.

\subsection{Comparison to 1313 \ce{La3Ni2O7}, trilayer \ce{La4Ni3O10} and \ce{Pr3Ni2O7}}

\begin{figure}[H]
    \centering
    \includegraphics[width=0.6\textwidth]{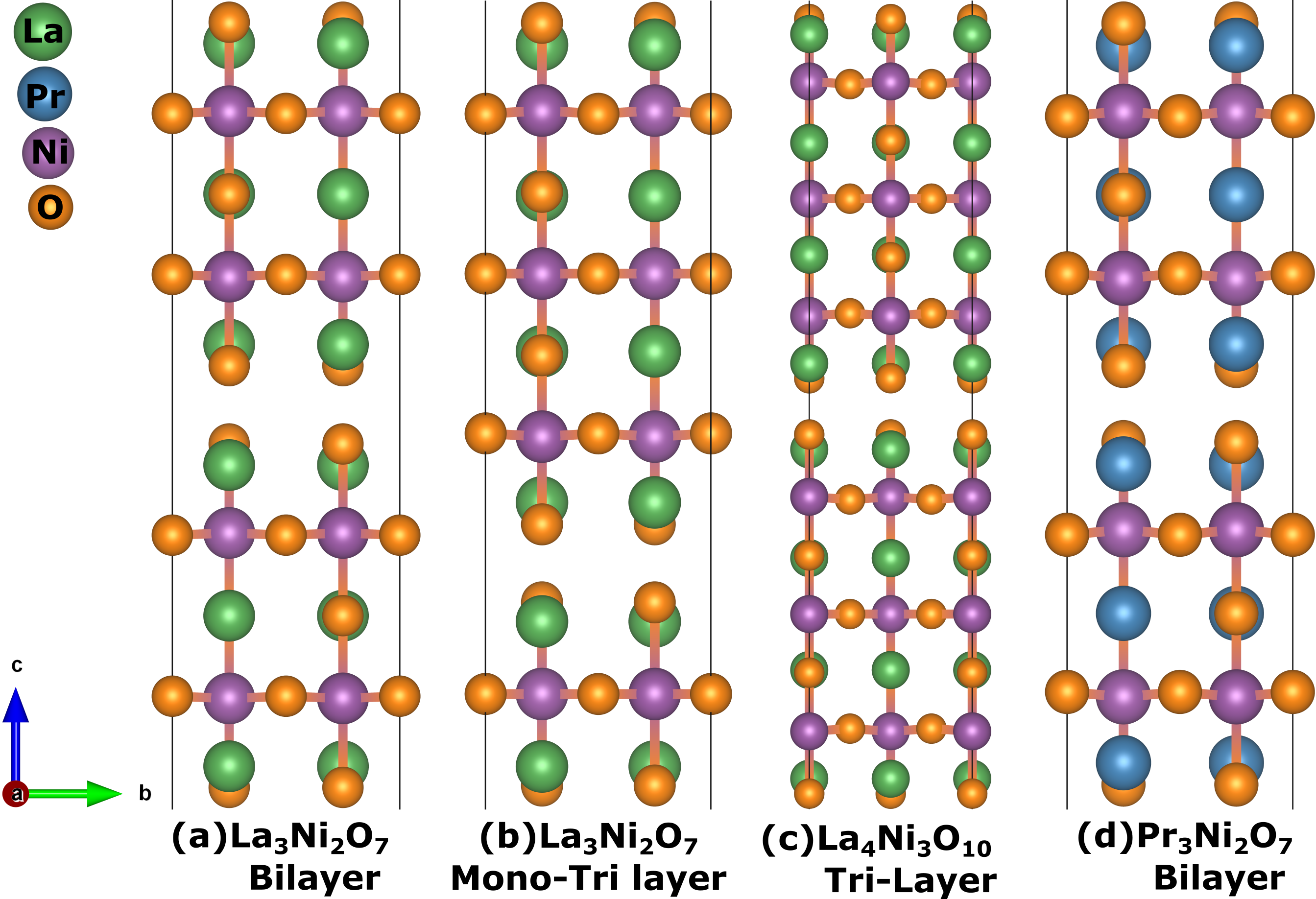} 
    \caption{ Structure of (a)bilayer $La_3Ni_2O_7$ (b) mono-layer tri-layer $La_3Ni_2O_7$ (c) tri-layer $La_4Ni_3O_{10}$ and (d) bilayer $Pr_3Ni_2O_7$}
    \label{fig:example6}
\end{figure}

Recent studies have revealed that \ce{La3Ni2O7} can exist in two distinct polymorphic forms, as illustrated in Figure~\ref{fig:example6}. The first is the well-established bilayer structure (Figure~\ref{fig:example6}a). The second is a less common monolayer–trilayer variant (Figure~\ref{fig:example6}b), which closely resembles the trilayer structure of \ce{La4Ni3O10} (Figure~\ref{fig:example6}c). In this monolayer–trilayer form of \ce{La3Ni2O7}, a Ni–O–Ni trilayer is interleaved with a monolayer, while in \ce{La4Ni3O10}, two such trilayers are separated by a rock-salt-type (La–O) layer. To examine how these structural motifs influence pressure-induced transitions, we performed calculations analogous to those described earlier for the bilayer phase, systematically exploring the effects of pressure and the on-site Coulomb interaction parameter \textit{U} on these polymorphs.

\begin{figure}[H]
    \centering
    \includegraphics[width=\textwidth]{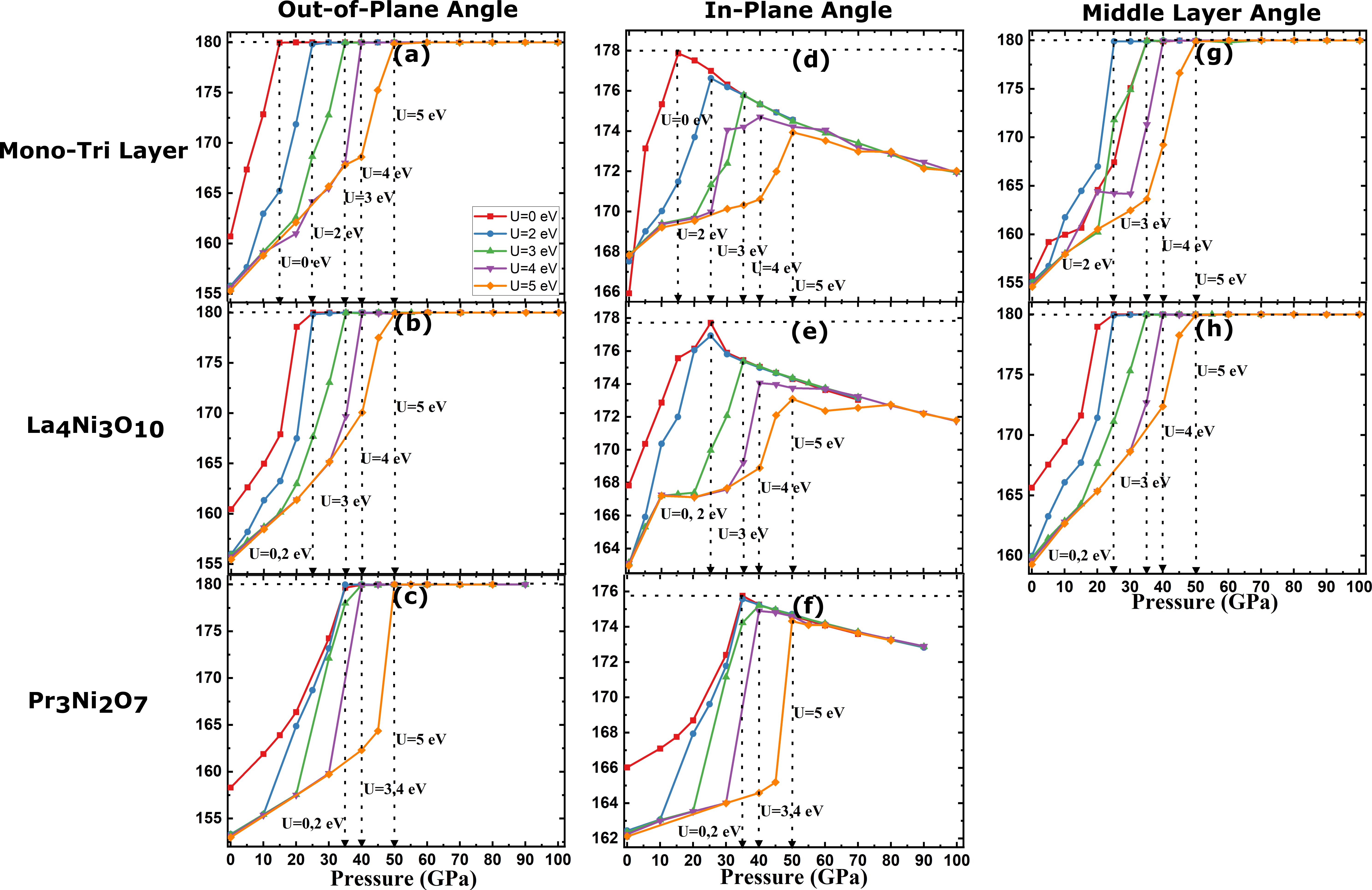} 
    \caption{(a), (b), (representts the effect on out-of-plane Ni-O-Ni bond angle on increasing \textit{U} and pressure on mono-layer tri-layer $La_3Ni_2O_7$, tri-layer $La_4Ni_3O_{10}$ and bilayer $Pr_3Ni_2O_7$ respectively. (d), (e) and (f) represent in-plane Ni-O-Ni variation of the same. (g) and (h) represents the middle layer in-plane Ni-O-Ni angle of  mono-layer tri-layer $La_3Ni_2O_7$, and tri-layer $La_4Ni_3O_{10}$.}
    \label{fig:example7}
\end{figure}

Figures~\ref{fig:example7}(a–h) summarize the evolution of key structural parameters as a function of pressure and \textit{U}. For the monolayer–trilayer \ce{La3Ni2O7}, the out-of-plane Ni–O–Ni bond angle increases with pressure (Figure~\ref{fig:example7}a). At \textit{U} = 0 eV, the transition pressure is $\sim$15 GPa, similar to that of the bilayer structure. However, the transition pressure increases significantly with increasing \textit{U}, reaching 25, 35, 40, and 50 GPa for \textit{U} = 2, 3, 4, and 5 eV, respectively. A comparable trend is observed for \ce{La4Ni3O10} (Figure~\ref{fig:example7}b). Notably, even at \textit{U} = 0 eV, its transition pressure is already higher ($\sim$25 GPa), aligning with that of the monolayer–trilayer \ce{La3Ni2O7} at \textit{U} = 2 eV. These findings suggest that the presence of additional Ni–O–Ni trilayers increases the structural rigidity and elevates the pressure required for bond straightening.

The in-plane Ni–O–Ni bond angles (Figures~\ref{fig:example7}d–e) also show dome-like evolution with pressure, a trend previously linked to superconductivity in bilayer nickelates. For both monolayer–trilayer \ce{La3Ni2O7} and trilayer \ce{La4Ni3O10}, the central Ni–O–Ni bond within each trilayer reaches 180$^\circ$ at the transition and remains flat at higher pressures (Figures~\ref{fig:example7}g–h), reflecting behavior analogous to the out-of-plane angle. At \textit{U} $\geq$ 3 eV, the structural response to pressure becomes nearly indistinguishable between these two systems, suggesting that the number of trilayer units is a critical factor in setting the transition pressure: a greater number of layers increases the structural resistance to straightening of the Ni–O–Ni bonds. This may arise from the steric constraint of displacing additional La and Ni atoms out of plane as the bond straightens, thus requiring higher pressure for structural reorganization.

To further investigate the role of rare-earth substitution and layer count, we replaced La with Pr in the bilayer \ce{La3Ni2O7} structure to form \ce{Pr3Ni2O7} (Figure~\ref{fig:example6}d). After structural optimization, we applied hydrostatic pressure while varying \textit{U}. As shown in Figure~\ref{fig:example7}c, \ce{Pr3Ni2O7} exhibits a strong increase in transition pressure with increasing \textit{U}, consistent with the behavior of La-based analogs. The slightly smaller ionic radius of Pr introduces greater lattice distortion at ambient pressure, leading to more bent Ni–O–Ni angles. Nevertheless, the in-plane Ni–O–Ni bond angle (Figure~\ref{fig:example7}f) retains its dome-like evolution with pressure, analogous to that in \ce{La3Ni2O7} and \ce{La4Ni3O10}.

Interestingly, recent experiments by Zhang et al.~\cite{zhang2025bulksuperconductivitypressurizedtrilayer} report a transition pressure of $\sim$80.1 GPa for \ce{Pr4Ni3O10}, significantly higher than that of \ce{Pr3Ni2O7}. This observation further supports the hypothesis that the transition pressure scales with the number of Ni–O–Ni layers. Together, these results provide the first theoretical insights into pressure-induced structural transitions in \ce{Pr3Ni2O7}, characterized by the flattening of out-of-plane Ni–O–Ni bond angles and a superconductivity-related dome in the in-plane angle. These findings underscore the critical interplay between layer architecture and rare-earth doping in governing structural transitions in layered nickelates under pressure.

\subsection{Compressive Strain Effect}

To assess the potential for strain-induced superconductivity in epitaxially grown thin films of \ce{La3Ni2O7}, we investigated the effect of biaxial compressive strain applied in the \textit{ab}-plane, while allowing full structural relaxation along the out-of-plane \textit{c}-axis. This strain condition simulates epitaxial growth on lattice-mismatched substrates, which can introduce substantial structural distortions, particularly in the in-plane Ni–O–Ni bond angles that are more susceptible to strain than their out-of-plane counterparts.

\begin{figure}[H]
    \centering
    \includegraphics[width=0.6\textwidth]{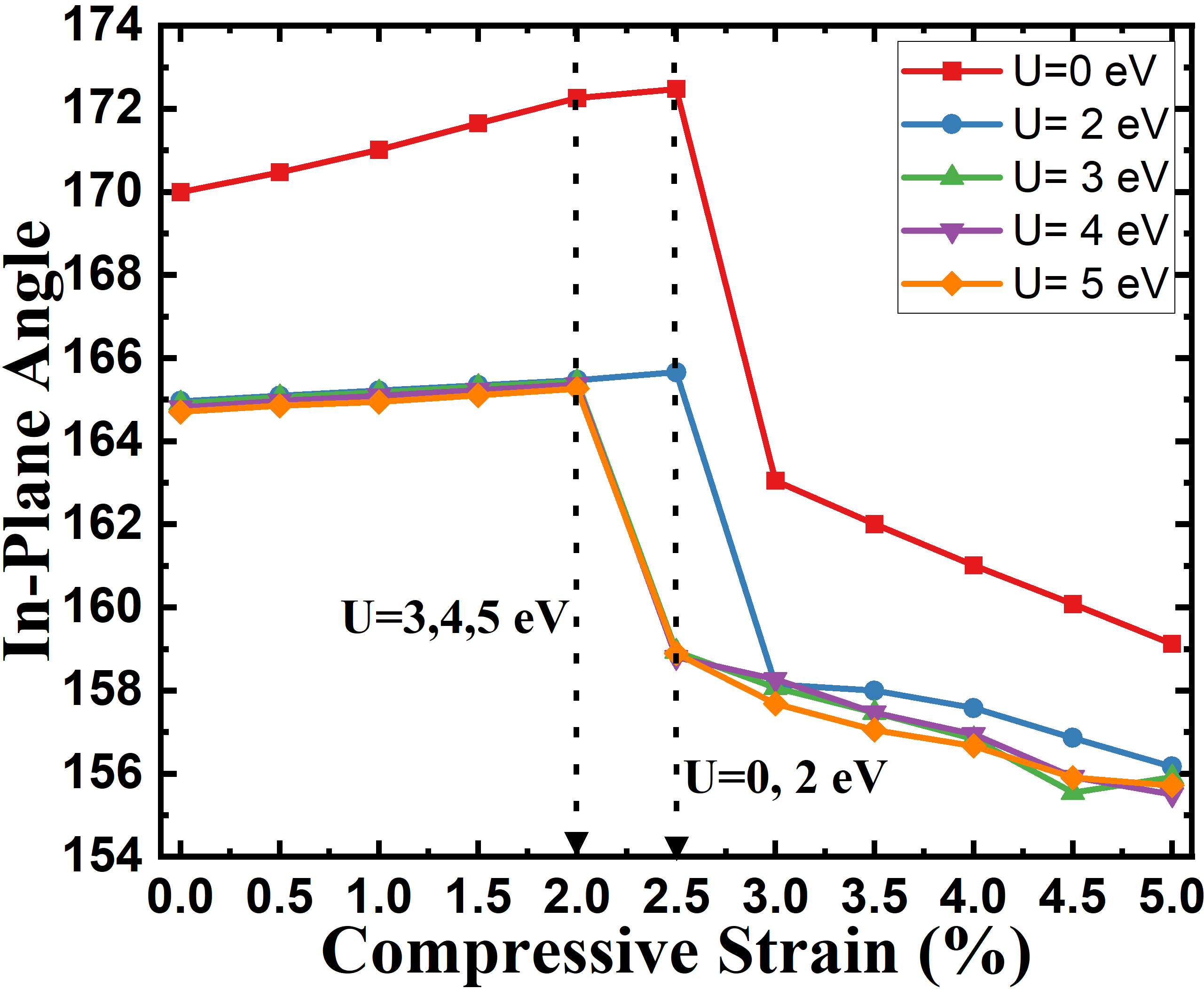} 
    \caption{In plane bond angles for La$_3$Ni$_2$O$_7$ versus compressive strain at different U. }
    \label{fig:example8}
\end{figure}

Figure~\ref{fig:example8} illustrates the evolution of the in-plane Ni–O–Ni bond angle as a function of compressive strain for various values of the on-site Coulomb interaction parameter \textit{U}. For \textit{U} = 0 and 2 eV, the bond angle increases with increasing compressive strain and reaches a maximum near 2.5\%. As \textit{U} increases to 3, 4, and 5 eV, this maximum shifts to lower strain values around 2.0\%. This non-monotonic behavior resembles the superconducting dome observed under hydrostatic pressure and suggests that similar optimal Ni–O–Ni geometries conducive to superconductivity can be achieved via epitaxial strain.

These results imply that the application of a moderate compressive strain (2–3\%)—achievable through careful substrate selection could stabilize structural conditions favorable for superconductivity in bilayer \ce{La3Ni2O7}, even in the absence of external pressure. This is in good agreement with the strain at which superconductivity was recently observed\citep{strain}. The bond angles, even at the optimum strain, are more bent than at the optimum pressure; in agreement with the hypothesis that in-plane bond angles strongly correlate with Tc, our results suggest a significantly lower $T_c$ than the highest one obtained from hydrostatic pressure.

We also investigated the influence of tensile strain, which corresponds to in-plane lattice expansion. In contrast to compressive strain, tensile strain induces a continuous reduction in both in-plane and out-of-plane Ni–O–Ni bond angles. This trend suggests that tensile strain disrupts the structural motifs associated with superconductivity and may suppress the onset of such phases.

Overall, our findings underscore the high tunability of the Ni–O–Ni bonding network in \ce{La3Ni2O7} through epitaxial strain. This tunability opens up a promising pathway to engineering superconducting phases in layered nickelates by optimizing the lattice geometry under ambient or mild thermodynamic conditions.

\subsection{Electronic Structure Evolution Under Pressure}
\begin{table}[h]
    \centering
    \begin{tabular}{lccc|ccc}
        \toprule
        \multicolumn{1}{c}{} & \multicolumn{3}{c}{\textbf{0 GPa}} & \multicolumn{3}{c}{\textbf{15 GPa}} \\
        \cmidrule(lr){2-4} \cmidrule(lr){5-7}
        \textbf{U = 2 eV} & \textbf{up} & \textbf{down} & \textbf{total} & \textbf{up} & \textbf{down} & \textbf{total} \\
        \midrule
        dz$^2$       & 0.95  & 0.40  & 1.35  & 0.87  & 0.57  & 1.44  \\
        dx$^2$-y$^2$ & 0.92  & 0.42  & 1.34  & 0.75  & 0.70  & 1.45  \\
        dzx         & 1.00  & 0.99  & 1.99  & 0.99  & 0.99  & 1.98  \\
        dzy         & 1.00  & 0.99  & 1.99  & 0.99  & 0.99  & 1.98  \\
        dxy         & 1.00  & 1.00  & 1.99  & 0.99  & 0.99  & 1.98  \\
        \bottomrule
    \end{tabular}
    \caption{Eigenvalues of Ni d orbitals in the FM  state of La$_3$Ni$_2$O$_7$ at 0 and 15 GPa for \textit{U}=2 eV.}
    \label{tab:occupancy}
\end{table}

To gain insight into the pressure-driven electronic evolution of ferromagnetic \ce{La3Ni2O7}, we computed the orbital occupancies and projected density of states (PDOS) of the Ni \textit{3d} manifold at ambient and high-pressure conditions, focusing on \textit{U} = 2 eV (Table~\ref{tab:occupancy}). As expected, the \( t_{2g} \) orbitals (\( d_{xy} \), \( d_{zx} \) and \( d_{yz} \)) are nearly fully occupied and lie well below the Fermi level. In contrast, the partially filled e$_g$ orbitals (\( d_{z^2} \) and \( d_{x^2-y^2} \)) dominate the electronic states near the Fermi level and play a central role in low-energy excitations. 

Upon applying pressure, we observe a redistribution of electronic density within the e$_g$ orbitals. The occupational eigenvalues of the \( d_{z^2} \) orbital decrease from 1.35 (higher energy) to 1.44 (lower energy), while that of the \( t_{2g} \) states remains essentially unchanged (from 1.99 to 1.98), indicating enhanced delocalization and hybridization of the e$_g$ electrons. These changes reflect the development of orbital polarization under pressure, which is known to influence electron transport and superconducting properties in correlated transition metal oxides.

\begin{figure}[H]
    \centering
    \includegraphics[width=0.9\textwidth]{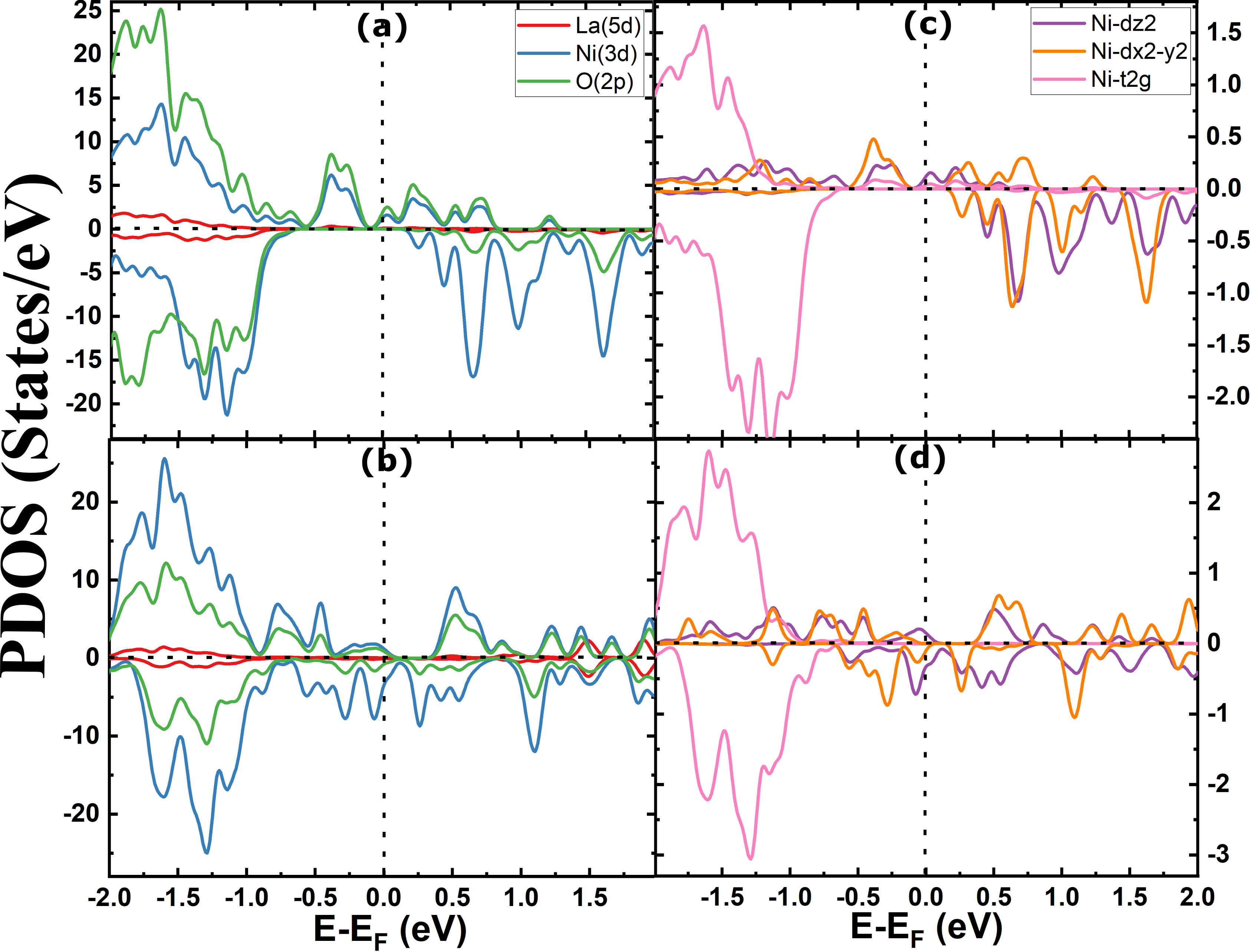} 
    \caption{FM La$_3$Ni$_2$O$_7$ PDOS plots. (a) and (b) show the PDOS at 0 and 15 Gpa for La(5d), Ni(3d), and O(2p). (c) and (d) Shows the Ni d-orbital PDOS at 0 and 15 GPa.}
    \label{fig:example9}
\end{figure}

PDOS analyses (Figure~\ref{fig:example9}) reveal that states near the Fermi level are primarily composed of Ni(3d) and O(2p) orbitals, with La (5d) states lying far below. At 0 GPa, the \( t_{2g} \) orbitals are localized below the Fermi level ($\sim$–1 eV), while the e$_g$ orbitals, particularly \( d_{z^2} \) lie at or near the Fermi level. Notably, the \( d_{x^2-y^2} \) states exhibit a pronounced bonding-antibonding splitting of $\sim$1  eV, consistent with strong hybridization due to the blocking layers.

\begin{figure}[H]
    \centering
    \includegraphics[width=0.9\textwidth]
    {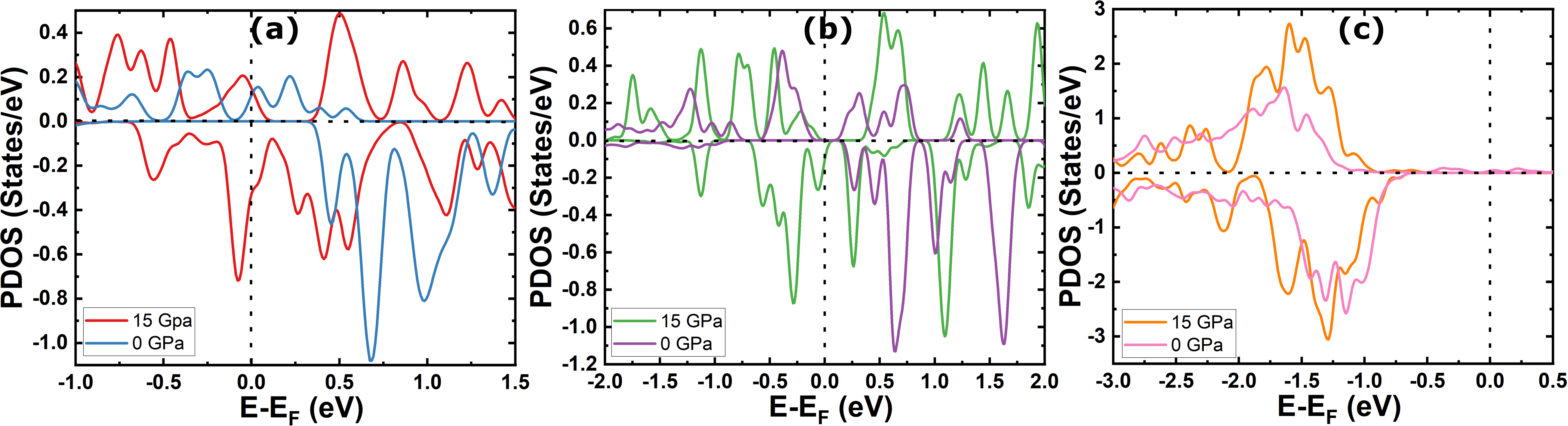} 
    \caption{FM La$_3$Ni$_2$O$_7$ PDOS plots comparison at 0 and 15 GPa. (a) \( d_{z^2} \) (b) \( d_{x^2-y^2} \) (c) \( t_{2g} \).}
    \label{fig:example10}
\end{figure}

Under 15 GPa pressure, the Ni (3d) states shift closer to the Fermi level (Figure~\ref{fig:example9}b), and both 
\( d_{z^2} \) and \( d_{x^2-y^2} \) orbitals exhibit increased spectral weight (Figure~\ref{fig:example10}), indicative of enhanced Ni–O covalency and greater orbital overlap. The \( d_{xz} \) and  \( d_{yz} \) states also become degenerate, suggesting improved symmetry and bandwidth under compression. The \( t_{2g} \) orbitals remain low in energy, suggesting minimal participation in bonding under pressure.

\begin{figure}[H]
    \centering
    \includegraphics[width=0.9\textwidth]{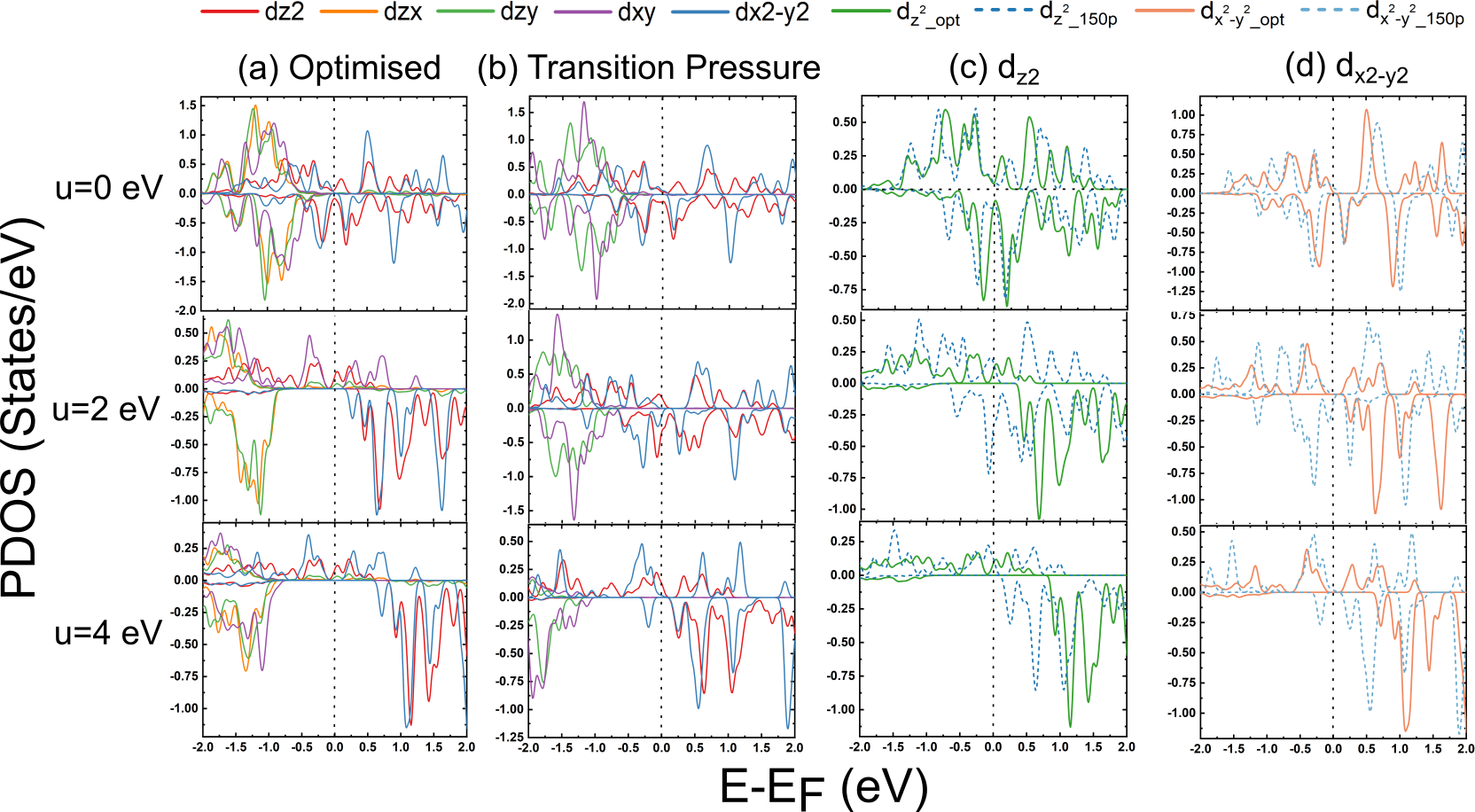} 
    \caption{FM La$_3$Ni$_2$O$_7$ PDOS plots. (a) Shows the d-orbital PDOS at 0 Gpa. (b) Shows the d-orbital PDOS at transition pressure. (c) dz$^2$ orbital at 0 and transition pressure. (d) dx$^2$-y$^2$ orbital at 0 and transition pressure.}
    \label{fig:example11}
\end{figure}

The influence of electron correlation is further evident in the evolution of the  e$_g$ PDOS with varying \textit{U} values (Figure~\ref{fig:example11}). At \textit{U} = 0 eV, both \( d_{z^2} \) and \( d_{x^2-y^2} \) orbitals retain significant weight at the Fermi level, favoring electronic delocalization. At \textit{U} = 2 eV, moderate bonding–antibonding splitting emerges, while at \textit{U} = 4 eV, the e$_g$ states are pushed further below the Fermi level, suppressing low-energy spectral weight and signaling enhanced localization.

\begin{figure}[H]
    \centering
    \includegraphics[width=0.9\textwidth]{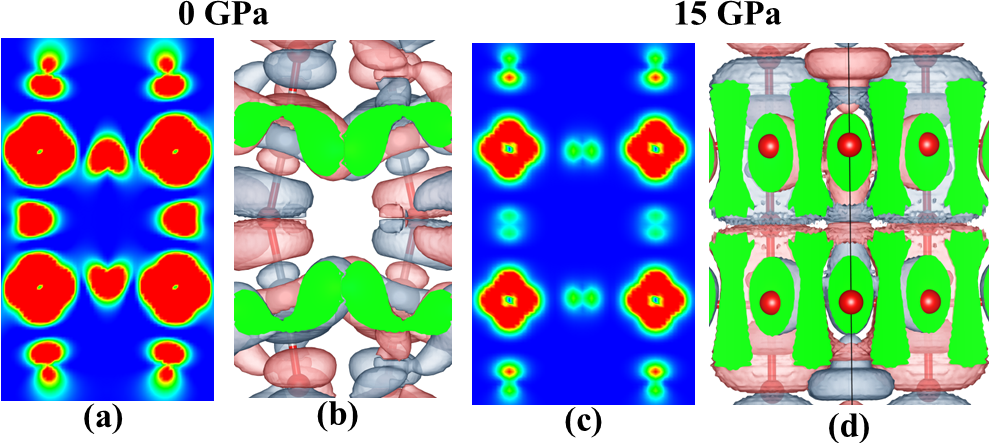} 
    \caption{(a) and (c) represent spin polarization while (b) and (d) represent an example of an electronic wavefunction at the Gamma point of \ce{La3Ni2O7}, represented at the same isovalue; blue and red correspond to different signs of the phase.}
    \label{fig:example12}
\end{figure}

To further visualize the evolution of bonding, we examined the spin-resolved charge densities and wavefunctions at the $\Gamma$-point (Figure~\ref{fig:example12}). At ambient pressure, the electron lobes (blue and red) are alternately aligned across Ni and O, indicative of antibonding character. Under 15 GPa pressure, the charge density becomes more continuous along the Ni–O–Ni bonds, especially in the \textit{ab}-plane, demonstrating increased overlap between Ni \( d_{x^2-y^2} \) and O(2p) orbitals. Along the \textit{c}-axis, we also observe enhanced \( d_{z^2} \)-O(2p) overlap, although the bonding remains comparatively weaker. 

Taken together, these results reveal that hydrostatic pressure induces a redistribution of spectral weight in favor of more delocalized e$_g$ states and stronger covalent bonding within the Ni–O network. The extent of this delocalization is modulated by the choice of \textit{U}, with moderate values (e.g., 2 eV) yielding a balance between itinerancy and correlation. Such an electronic environment is likely conducive to the emergence of superconductivity, echoing trends observed in other correlated oxide systems.

\section{Conclusion}

We have systematically investigated the pressure and strain induced structural and electronic evolution of layered nickelates using density functional theory with Hubbard \textit{U} corrections (DFT+\textit{U}). We found the Ni-O-Ni in plane bond angles versus pressure strongly correlate with the $T_c$, while versus strain we found that the bond angles are straightest at 2\% compressive strain - the strain at which superconductivity was observed at ambient pressure. Analyzing lattice parameters, Ni–O bond lengths, and Ni–O–Ni bond angles over a range of pressures and \textit{U} values, we capture the structural transition in \ce{La3Ni2O7} from an orthorhombic to a high-symmetry tetragonal phase characterized by the straightening of the out-of-plane Ni–O–Ni bond angle to 180$^\circ$ in agreement with experimental reports, coinciding with the suppression of NiO$_6$ octahedral tilting. Replacement of La with heavier Lanthanides (e.g. Pr) increases the transition pressure, and so does increasing the number of perovskite layers.

Our results reveal that the critical transition pressure increases with \textit{U}, reflecting a competition between electron correlation and external pressure. This interplay also governs the stability of the spin states: at lower \textit{U} (notably at \textit{U} = 2 eV), a high-spin to low-spin transition emerges near the superconducting pressure regime, whereas higher \textit{U} values stabilize the high-spin state across the full pressure range, shifting the superconducting window to higher pressures. This suggests that while correlations are known to be important to superconductivity in these materials, a just-right amount of correlation is key: correlations that are too strong will prevent the bond angles from straightening.

Electronic structure analysis shows that the Ni $e_g$ orbitals dominate the density of states near the Fermi level. Under compression, these orbitals shift downward in energy, enhancing Ni–O hybridization and promoting electronic delocalization. Additionally, we find that the number of Ni–O–Ni interlayer linkages plays a key role in determining the critical pressure, emphasizing the importance of lattice connectivity in tuning superconducting behavior.

In summary, this work elucidates the competing roles of pressure and electronic correlations in governing the structural and electronic landscape of layered nickelates. These insights provide a foundation for guiding future experimental and theoretical strategies to engineer superconductivity through strain, pressure, and chemical doping in nickelate systems.

\section{Computational Methods}
We have used the first principles density functional theory (DFT) calculations to understand the effect of pressure and electron correlation on the superconductivity of \ce{La3Ni2O7 } and its family members. For all calculations, we have used the DFT QUANTUM ESPRESSO code \cite{giannozzi2009quantum} along with the Vanderbilt Ultra-Soft Pseudopotential developed by David Vanderbilt \cite{vanderbilt1990ultrasoft}. A cut-off energy of 40 Ry was fixed for the plane wave expansion of the wave function. An automatic sampling of the reciprocal space with a 4x4x2 K-mesh was employed for all calculations. The force and energy convergence thresholds of 0.01 and $10^{-4}$ eV were applied, respectively. The Gaussian smearing was set to an energy width of 0.004 Ry. The initial structure of \ce{La3Ni2O7 } is taken from the Materials Project \cite{materialsproject18926}. We have then fully relaxed the lattice parameter and the ions to obtain a ground state. With this ground state, we have applied pressure from 5 GPa to 100 GPa and compressive strain from 0\% \text{ to } 5\% using the \texttt{cell\_dofree} parameter and relaxing in the z direction. At first, we used the ferromagnetic (FM) structure with spin, with no electron correlation correction. After that, we varied the Hubbard U parameter from 0 to 5 eV and calculated the effect of it on the transition pressure and structure. To study the effect of magnetic ordering, we have used three antiferromagnetic (AFM) structures, namely, A-AFM, C-AFM, and G-AFM, as discussed in previous sections.

\begin{acknowledgement}

B.S and A.B.G would like to acknowledge Indiana University startup funds. This work is supported by the Lilly Endowment, Inc., through its support for the Indiana University Pervasive Technology Institute, which allowed DFT computations to be performed at Indiana University on the BigRed200 and Quartz high-performance computing clusters.

\end{acknowledgement}


\bibliography{acs-achemso}

\end{document}